\documentclass[11pt,a4paper]{article}
\usepackage{amssymb} \usepackage{amsmath} \usepackage{graphicx}
\usepackage{epsfig,latexsym}
\begin{document}

\def\bef{\begin{figure}}
\def\eef{\end{figure}}
\newcommand{\ans}{ansatz }
\newcommand{\be}[1]{\begin{equation}\label{#1}}
\newcommand{\beq}{\begin{equation}}
\newcommand{\ee}{\end{equation}}
\newcommand{\beqn}[1]{\begin{eqnarray}\label{#1}}
\newcommand{\eeqn}{\end{eqnarray}}
\newcommand{\bd}{\begin{displaymath}}
\newcommand{\ed}{\end{displaymath}}
\newcommand{\mat}[4]{\left(\begin{array}{cc}{#1}&{#2}\\{#3}&{#4}
\end{array}\right)}
\newcommand{\matr}[9]{\left(\begin{array}{ccc}{#1}&{#2}&{#3}\\
{#4}&{#5}&{#6}\\{#7}&{#8}&{#9}\end{array}\right)}
\newcommand{\matrr}[6]{\left(\begin{array}{cc}{#1}&{#2}\\
{#3}&{#4}\\{#5}&{#6}\end{array}\right)}
\newcommand{\cvb}[3]{#1^{#2}_{#3}}
\def\lsim{\raise0.3ex\hbox{$\;<$\kern-0.75em\raise-1.1ex
e\hbox{$\sim\;$}}}
\def\gsim{\raise0.3ex\hbox{$\;>$\kern-0.75em\raise-1.1ex
\hbox{$\sim\;$}}}
\def\abs#1{\left| #1\right|}
\def\simlt{\mathrel{\lower2.5pt\vbox{\lineskip=0pt\baselineskip=0pt
           \hbox{$<$}\hbox{$\sim$}}}}
\def\simgt{\mathrel{\lower2.5pt\vbox{\lineskip=0pt\baselineskip=0pt
           \hbox{$>$}\hbox{$\sim$}}}}
\def\unity{{\hbox{1\kern-.8mm l}}}
\newcommand{\eps}{\varepsilon}
\def\ep{\epsilon}
\def\ga{\gamma}
\def\Ga{\Gamma}
\def\om{\omega}
\def\omp{{\omega^\prime}}
\def\Om{\Omega}
\def\la{\lambda}
\def\La{\Lambda}
\def\al{\alpha}
\newcommand{\ov}{\overline}
\renewcommand{\to}{\rightarrow}
\renewcommand{\vec}[1]{\mathbf{#1}}
\newcommand{\vect}[1]{\mbox{\boldmath$#1$}}
\def\tm{{\widetilde{m}}}
\def\mcirc{{\stackrel{o}{m}}}
\newcommand{\Dm}{\Delta m}
\newcommand{\dm}{\varepsilon}
\newcommand{\tanb}{\tan\beta}
\newcommand{\nbar}{\tilde{n}}
\newcommand\PM[1]{\begin{pmatrix}#1\end{pmatrix}}
\newcommand{\up}{\uparrow}
\newcommand{\down}{\downarrow}
\def\omE{\omega_{\rm Ter}}
%

\newcommand{\Dsusy}{{susy \hspace{-9.4pt} \slash}\;}
\newcommand{\DCP}{{CP \hspace{-7.4pt} \slash}\;}
\newcommand{\mc}{\mathcal}
\newcommand{\gr}{\mathbf}
\renewcommand{\to}{\rightarrow}
\newcommand{\gtc}{\mathfrak}
\newcommand{\wh}{\widehat}
\newcommand{\br}{\langle}
\newcommand{\kt}{\rangle}


\def\lsim{\mathrel{\mathop  {\hbox{\lower0.5ex\hbox{$\sim$}
\kern-0.8em\lower-0.7ex\hbox{$<$}}}}}
\def\gsim{\mathrel{\mathop  {\hbox{\lower0.5ex\hbox{$\sim$}
\kern-0.8em\lower-0.7ex\hbox{$>$}}}}}

\def\nn{\\  \nonumber}
\def\de{\partial}
\def\brf{{\mathbf f}}
\def\bbf{\bar{\bf f}}
\def\bF{{\bf F}}
\def\bbF{\bar{\bf F}}
\def\bA{{\mathbf A}}
\def\bB{{\mathbf B}}
\def\bG{{\mathbf G}}
\def\bI{{\mathbf I}}
\def\bM{{\mathbf M}}
\def\bY{{\mathbf Y}}
\def\bX{{\mathbf X}}
\def\bS{{\mathbf S}}
\def\bb{{\mathbf b}}
\def\bh{{\mathbf h}}
\def\bg{{\mathbf g}}
\def\bla{{\mathbf \la}}
\def\bmu{\mathbf m }
\def\by{{\mathbf y}}
\def\bmu{\mbox{\boldmath $\mu$} }
\def\bsig{\mbox{\boldmath $\sigma$} }
\def\bunity{{\mathbf 1}}
\def\cA{{\cal A}}
\def\cB{{\cal B}}
\def\cC{{\cal C}}
\def\cD{{\cal D}}
\def\cF{{\cal F}}
\def\cG{{\cal G}}
\def\cH{{\cal H}}
\def\cI{{\cal I}}
\def\cL{{\cal L}}
\def\cN{{\cal N}}
\def\cM{{\cal M}}
\def\cO{{\cal O}}
\def\cR{{\cal R}}
\def\cS{{\cal S}}
\def\cT{{\cal T}}
\def\eV{{\rm eV}}
%





\large
 \begin{center}
 {\Large \bf Dynamical R-parity violations from exotic instantons}
 \end{center}

 \vspace{0.1cm}

 \vspace{0.1cm}
 \begin{center}
{\large Andrea Addazi}\footnote{E-mail: \,  andrea.addazi@infn.lngs.it} \\
{\it \it Dipartimento di Fisica,
 Universit\`a di L'Aquila, 67010 Coppito, AQ \\
LNGS, Laboratori Nazionali del Gran Sasso, 67010 Assergi AQ, Italy}
\end{center}

\vspace{1cm}
\begin{abstract}
\large
We show how R-parity can be dynamically broken 
by non-perturbative quantum gravity effects. 
In particular, in D-brane models, 
Exotic instantons provide a simple and 
calculable mechanism for the generation of R-parity violating
bilinear, trilinear and higher order superpotential terms.  
We show examples of MSSM-like D-brane models, 
in which one Exotic Instanton induces 
only one term among the possible R-parity violating superpotentials. 
Naturally, the idea can be generalized for other gauge groups. 
 As a consequence, a dynamical violation 
of R-parity does not necessarily destabilize the proton,
{\it i.e.} a strong fine tuning is naturally avoided, in our case. 
For example, a Lepton violating superpotential term can be generated 
without generating Baryon violating terms, and {\it viceversa}. 
This has strong implications in phenomenology:
neutrino, neutron-antineutron, electric dipole moments, dark matter and LHC physics. 

\end{abstract}

\baselineskip = 20pt

\section{Introduction}

The possibility, that MSSM
does not possess an R-parity
has intriguing implications for phenomenology,
in particular for LHC, baryon/lepton violations 
in low energy physics, neutrino mass and so on.
This subject is rich 
with reviews and papers.
See 
 \cite{R1,R2,R3,R5,R6,R7,R8,R9,R10,R11,R12,R13,R14,R15,R16,R17,R18,R19,R20},
for a general overview in R-violating models.

However, as it is well known, MSSM without R-parity 
immediately  destabilizes the proton,
as well as the lightest neutralino. 
For the neutralino, one  need not be particularly afraid:
maybe, it could be substituted by another candidate.
For example, gravitino 
is an alternative 
candidate for dark matter. 
 On the other hand,
proton destabilization
is a serious problem:
MSSM, without extra discrete symmetries,
has to assume a very strong fine-tuning. 
For this reason, such a proposal seems 
farfetched without a deeper theoretical reason. 
On the other, we have learned several times in particle physics 
that, often, a mechanism of spontaneous or dynamical breaking 
of a symmetry is smarter than an explicit one. 
Can R-parity be spontaneously or dynamically broken, without proton destabilization?

As shown in \cite{Ibanez1,Ibanez2,Ibanez3,Florea,Blu1,Blu2,Cvetic1,Cvetic2,Cvetic3,Koba1,Koba2,Addazi1,Addazi2,Addazi3,Addazi4,Addazi5,Addazi7},
 R-parity can be dynamically broken 
by {\it exotic instantons}. 
In particular, in intersecting D-brane models 
with open strings attached to D-brane stacks, 
exotic instantons are nothing but other Euclidean D-branes (or E-branes)
wrapping differently the 
Calabi-Yau compactification \footnote{However, another class of exotic instantons studied in \cite{Parsa1,Parsa2,Parsa3} could lead to the same relevant effects. I would like to thank Parsa Ghorbani for useful discussions of these aspects.  }  (different n-cycles) with respect 
to physical D-branes (see \cite{DMSSM1,DMSSM2,DMSSM3,DMSSM4,DMSSM5,DMSSM6,DMSSM8,DMSSM9,DMSSM10,DMSSM11,DMSSM12,DMSSM13,DMSSM14,DMSSM15,Jim1,Jim2,Jim3,Jim4} for D-brane models 
reproducing MSSM in the low energy limit)
\footnote{An alternative mechanism for a dynamical R-parity 
violation is considered in \cite{Csaki1,Csaki2}. In this one, 
R-parity breaking is communicated from a hidden sector
to our ordinary one. 
In this case, R-violating K$\ddot{a}$hler potentials 
are generated. On the other hand, another simple mechanism 
for a spontaneous R-parity breaking was proposed in \cite{Perez1,Perez2,Perez3}. 
This last seems intriguingly connected with our suggestion: 
usually, exotic instantons' effects are connected to a Stueckelberg
mechanism for $U(1)_{B-L}$, as shown in publications cited above. 
} \footnote{Let us comment that $SU(5)$ models
can be embedded in D-brane models and 
that also in this case exotic instantons can generate R-parity violating
terms. This can be an interesting reinterpretation 
of models like the one considered in \cite{GUT1}.
Alternatively, one can construct 3-3-1 models, like the one in
\cite{Valle1,Valle2,Valle3}, from D-branes constructions,
in which exotic instantons generate 
extra B/L-violating effective operators,
not permitted at perturbative level.
I would like to thank Luca Di Luzio and Jos\'e Valle
for inspiring conversations on these subjects.}. 

In this paper, we discuss several different implications 
in phenomenology of exotic instantons, in the 
particular case of MSSM.
We will show how, starting from an R-parity preserving model, 
one can generate specific operators in the superpotentials
from exotic instantons, without generating all the possible 
R-violating violating terms! 
This leads to very interesting 
implications for LHC physics, Neutrino Physics, Dark Matter issues,
Neutron-Antineutron physics, Electric Dipole moment physics, without proton decay.

\section{Dynamical generation of bilinear and trilinear superpotential terms}

As first examples, let us consider a D-branes' model as the one in Fig. 1-(a)-(b)-(c). 
At low energy limit, these reproduce 
$\mathcal{N}=1$ susy 
$G=U(3)_{c}\times Sp(2)_{L}\times U(1) \times U'(1)\times U''(1)$,
embedding MSSM \footnote{$U(1)_{Y}=-\frac{1}{3}U(1)_{3}+U(1)+U(1)'-U(1)''$ for Fig.1-(c). However, 
following considerations are valid for a more general class of models with different
hypercharge combinations. }. We consider a
 $\Omega$-plane in our construction.
Let us remind that: i) extra anomalous $U(1)$s 
contained in $G$ are cured by 
the Generalized Chern-Simon mechanism, 
in string theory;
ii) extra $Z'$ bosons associated to extra $U(1)$s 
get masses through a Stuckelberg mechanism \footnote{
We mention that another intriguing application of Stuckelberg mechanism is considered in 
Massive gravity. For a study of geodetic instabilities,
for a class of these models, see \cite{Addazi:2014mga}.},
typically $m_{Z'}\sim M_{S}$, where $M_{S}$ is the 
string scale. These aspects are extensively discussed 
in \cite{GCS1,GCS2,GCS3,GCS4,GCS5,GCS6,GCS7,GCS8,GCS9,GCS10,GCS11}. 
However, the presence of Euclidean D2-branes  (usually also called E2-instantons), 
wrapping different 3-cycles with respect to ordinary D6-branes 
stacks, induce extra superpotential terms 
not permitted at perturbative level by R-parity. 
Which superpotential operators?
It depends on intersections of the $E2$-instanton 
with ordinary D6-branes.
In particular, in cases of Fig.1-(a)-(b), $E2$-instantons 
have a $O(1)$ Chan-Paton group \footnote{Such an $E2$-instanton
has to stay on a $\Omega^{+}$-plane, 
while for ordinary $D6$-branes in Fig.1-(a)-(b)
are projected by an $\Omega^{-}$-plane.
So, our $\Omega$-plane in Fig.1-(a)-(b)
"switches" from $\Omega^{+}$ to $\Omega^{-}$,
compatible with our quiver. }.
In Fig.1-(a), we consider an E2-instanton intersecting one time 
with the $U(1)$-stack, one time with $U'(1)$-stack, 
one time with $Sp_{L}(2)$-stack \footnote{
 $\Omega$-planes are introduced for cancellations 
 of stringy tadpoles. They are important for the construction of realistic models of particle physics from open string theories
 \cite{Sagnotti1,Sagnotti2,Sagnotti3,Sagnotti6,Sagnotti7,Sagnotti8,Bianchi:1990yu, Bianchi:1990tb, Bianchi:1991eu, sessantatre, sessantaquattro, MBJFM, Bianchi:1990yu, Bianchi:1990tb, Bianchi:1991eu, Angelantonj:1996uy, Angelantonj:1996mw}.}. 
Now, although similar calculations were made several times in the literature cited above, we
will show them again for our case, for completeness.
The following interactions are generated in Fig.1-(a):
\be{HL}
\mathcal{L}_{1}\sim C^{(1)}\beta^{(1)} H_{u_{A}}\tau_{A}^{(1)} +C_{i}^{'(1)}\gamma^{(1)} L_{A}^{i}\tau_{A}^{(1)}
\ee
where $\beta^{(1)},\gamma^{(1)},\tau^{(1)}$ are fermionic zero modes 
corresponding to excitations of open strings attached
to $U(1)-E2$, $U'(1)-E2$ and $Sp_{L}(2)-E2$ respectively; 
$i,j$ are $Sp_{L}(2)$ indices, $A;B$ are flavor indices;
$C^{(1)},C^{'(1)}$ are coupling constants, coming from the disk correlators.
Integrating out fermionic zero modes, 
we obtain 
\be{integration}
\mathcal{W}_{1}=\int d^{2}\tau^{(1)} d\beta^{(1)} d\gamma^{(1)} e^{\mathcal{L}_{1}}=M_{S}e^{-S_{E2}}(C^{(1)}C^{'(1)}_{i}) H_{u}L^{i}  
\ee
where $M_{S}$ is the string scale, $e^{-S_{E2}}$ depends 
on geometric moduli parametrizing 
3-cycles, wrapped by the $E2$-instanton on the $CY_{3}$.  
As a consequence, a R-parity violating superpotential 
$\mu'_{i}H_{u}L^{i}$ is generated by $E2$-instanton 
in Fig.1-(a), with $\mu'_{i}=M_{S}e^{-S_{E2}}(C^{(1)}C^{'(1)}_{i})$. Note that $e^{-S_{E2}}$ can be in principle 
$e^{-S_{E2}}\sim 1$ as well as $e^{-S_{E2}}\sim 10^{-20}$:
this depends on the particular geometry of 3-cycles
wrapped by $E2$-brane. The first case corresponds 
to small radii of 3-cycles, the second case to very large 
ones. From an effective theory point of view, 
 $\mu'$ can be assumed as a free-parameter,
attending for a realistic completion of this model. 
Now, let us consider another case, 
shown in Fig.1-(b),
with a different $E2$-instanton, 
that we call $E2'$. In fact, intersections of $E2'$-brane 
with stacks are very different with respect to the previous case.  
In this case, we consider $U(3)-E2'$, $U(1)-E2'$, $U'(1)-E2'$
intersections. In particular, we can consider $E2$ intersecting 
two times $U(3)$, one times $U(1)$ and $U(1)'$, with 
orientations shown in Fig.1-(b). 
Effective interactions between fermionic modulini and ordinary fields are 
\be{UDD}
\mathcal{L}_{2}\sim C^{(2)}_{i}\beta^{(2)}U^{c^{i}}_{A}\tau_{A}^{(2)} + C^{'(2)}_{j}\gamma^{(2)}D^{c^{j}}_{B}\tau_{B}^{(2)}
\ee
and integrating out over modulini space we obtain
\be{integration2}
\mathcal{W}_{2}=\lambda''U^{c}D^{c}D^{c}
\ee
where $\lambda''_{ijk,ABC}=(C^{(2)}C'^{(2)}C'^{(2)})_{ijk}e^{-S_{E2'}}\epsilon_{ABC}$. 
\begin{figure}[t]
\centerline{ \includegraphics [height=9cm,width=1.3 \columnwidth]{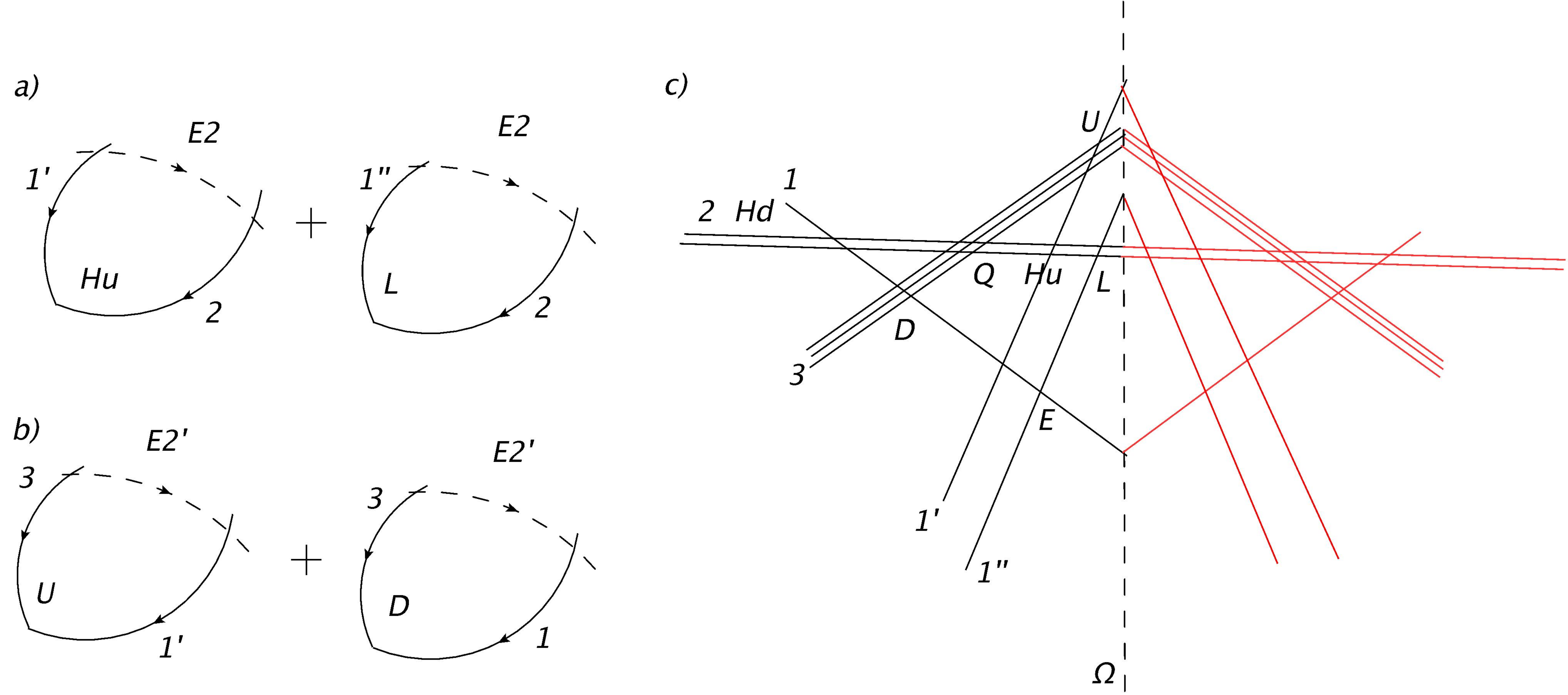}}
\vspace*{-1ex}
\caption{a) Mixed Disk amplitude generating a bilinear term $\mu' H_{u}L$.
b) Mixed Disk amplitude generating a trilinear term $\lambda'' U^{c}D^{c}D^{c}$. Notation in Figures: $U\equiv U^{c}$, $D\equiv D^{c}$, $E\equiv E^{c}$; $1,2,3$ are stacks of one, two and three parallel 
D6-branes, with 3-cycles on the Calabi-Yau $CY_{3}$;
in red the image 
of the D-branes system with respect to the $\Omega$-plane. }
\label{plot}   
\end{figure}
Similarly, with other D-branes' models, we can produce 
other possibile trilinear operators 
from $E2$-branes with the appropriate intersections
with ordinary $D6$-branes. 
In particular, in R-violating MSSM, 
we can produce the following superpotentials 
\be{RPV}
\mathcal{W}_{RPV}=\mathcal{W}_{1}+\mathcal{W}_{2}+\mathcal{W}_{3}+\mathcal{W}_{4}
\ee
with 
\be{RPV1}
\mathcal{W}_{1}=\mu'_{i}L_{i}H_{u}
\ee
\be{RPV2}
\mathcal{W}_{2}=\lambda''_{ijk}U_{i}^{c}D_{j}^{c}D_{k}^{c}
\ee
\be{RPV3}
\mathcal{W}_{3}=\lambda_{ijk}L_{i}L_{j}E^{c}_{k}
\ee
\be{RPV4}
\mathcal{W}_{4}=\lambda'_{ijk}Q_{i}L_{j}D_{k}^{c}
\ee
where $i,j,k=1,2,3$ are generation indices.
One can find 
that 
$\lambda,\lambda'\sim e^{-S_{E2'',E2'''}}$,
in generic MSSM D-brane models. 

Let me conclude this section remarking that
exotic instantons can be  "not democratic" with flavors.
In other words, mixed disk amplitudes like the ones
 generating (\ref{HL})-(\ref{UDD}),
can have 
{\it i.e} matrices $C,C'$, parametrizing flavor hierarchies, 
easily reaching splitting of $10^{1\div 3}$ orders among different generations. 
Such a situation is possible if:
i) 3-cycles (of $E2$-instantons or of $D6$-branes) have the same homologies 
but they are not identical ones; 
ii) 3-cycles have different homologies \footnote{I would like to thank Massimo Bianchi for useful comments on these aspects.}.
These aspects will have intriguing consequences for phenomenology
as we will see in the next sections.

\section{Phenomenology}

With respect to explicit R-parity violations of MSSM,
we would like to remark two important
aspects for phenomenology:
i) $\lambda,\lambda',\lambda''$ appear with factors $e^{-S_{E2}}$, 
geometrically understood as $E2$-brane wrapping 3-cycles on $CY_{3}$.
$e^{-S_{E2}}$ can be $<<1$ (large 3-cycles), or $\sim 1$ (small 3-cycles). 
ii) $E2$-brane of Fig.1-(a) generates one and only one R-parity violating superpotential term (\ref{RPV1}) among all the possible bilinear and trilinear terms. On the other hand, $E2'$-brane of Fig.1-(b) generates only (\ref{RPV2}). In explicitly R-violating MSSM
one has to consider in principle all superpotential terms, 
not avoided by R-parity. On the contrary, a dynamical breaking generates only 
one or at least a subclass of all the possible superpotential terms! 

Because of a situation with more $E2$-instantons complicate 
D-brane constructions, in our scenario, a model with one and only one 
superpotential term seems simpler than another one with all 
possible terms in (\ref{RPV}). As a consequence, 
our point of view is "inverted" with respect a model 
without R-parity: a situation with all superpotential 
terms is more complicated to be obtained, in our case. 

Finally, we would like to comment the rule of supersymmetry in these
mechanism. In fact, non-renormalization theorem guarantees 
that also after R-parity breaking, other R-parity violating superpotential 
cannot be generated by quantum corrections. 
On the other hand, possible non-holomorphic terms in the 
K$\ddot{a}$hler potential, generated at quantum level, may be relevant in our analysis.   
In particular, extra R-parity violating terms can lead to
proton decay. Unfortunately, to calculate such corrections 
is a difficult technical problem in a realistic D-branes model.
However, as commented in \cite{Addazi2}, 
one can reasonable assume that such corrections will be 
absent or at least negligible, in a class of models as the one considered 
here. 

\subsection{Neutrino Physics}
Let us discuss phenomenological implications 
of the mixed disk amplitude in Fig.1-(a), generating only one 
R-parity violating term $\mu' H_{u}L$.
In this case, Lepton number is violated,
but proton is not destabilized by any other superpotential 
terms, as well as no-baryon violating processes are generated. 
In this case, neutrini-neutralini mixings are induced 
by Sneutrino VEVs. This leads to a see-saw mechanism,
giving, at three level, a mass to one neutrino;
while the second neutrino mass scale
is generated by loop corrections
\cite{17,18}.
The same VEVs enter in the lightest neutralino decays 
$\tilde{\chi}_{1}^{0}\rightarrow \mu jj$ \cite{20},
and chargini decays $\tilde{\chi}_{1}^{0}\rightarrow lll,\tau ll, l \bar{b}b, \tau b\bar{b}$ \cite{30a,30b,30c}. In our model, 
this scenario corresponds to large 3-cycles 
of the Exotic Instanton involved, 
{\it i.e} $e^{-S_{E2}}<<1$,
assuming string scale as $M_{S}\simeq 10^{19}\, \rm GeV$. 
For a recent discussion of implications for LHC, see \cite{M1}.
This case is particularly interesting also because 
there are not other insidious bounds
from low energy physics and cosmology.
For instance, sphalerons and the bilinear term not wash-out 
all the initial Baryon/Lepton asymmetry \cite{Cheriguene:2014bxa}.
On the other hand, present bounds on CP-violating phases 
from electric dipole moments
are not in contradiction with cosmological ones \cite{Cheriguene:2014bxa}. 
We also note that in this case, 
neutralino cannot be a stable WIMP.
However, gravitino remains a good candidate for dark matter
with good relic abundance.
In RPV-models, gravitino 
can decay into $\tilde{G}\rightarrow \gamma \nu,Z\nu,Wl,h\nu$
(depending on its mass), with possible implications in
indirect detection of dark matter.
See \cite{Gravitino1,Gravitino2,Gravitino3,Gravitino4,Gravitino5,Gravitino6,Gravitino7,Gravitino8,Gravitino9,Gravitino10} for several papers in gravitino dark matter without R-parity.

Another possibility is to consider a model with two $E2$-instantons
generating $\mu' H_{u}L$ and $\lambda'QLD^{c}$. 
This case is also more intriguing:
Majorana masses for neutrini can be generated 
by squarks-quark radiative corrections, with intriguing signatures 
for LHC and $0\nu\beta\beta$-decay \cite{A1,A2,A3,A4,A5,Al1,Al2}. 
However, this case has more insidious bounds to avoid:
i) cosmological bounds for Baryon Asymmetry in our Universe (see Appendix A); ii) contrains from mesons physics, 
and in particular from tree-level $K-\bar{K}$, $B_{d}-\bar{B}_{d}$ and $B_{s}-\bar{B}_{s}$ mixings \cite{KS}.
As regards the cosmological bound,
we will discuss this one in Subsection 3.4.
As regards tree-level $B_{d}-\bar{B}_{d}$ oscillations, 
$\lambda'_{i1(2)3},\lambda'_{i31(2)}$
are restricted up to $8 \times 10^{-8} (m_{\tilde{l}}/100\, \rm GeV)^{2}$, approximately corresponding to 
$10^{-6}$ for $300\, \rm GeV$ \cite{KS}. 
As regards $K-\bar{K}$, one can get for 
$|\lambda'_{i12}\lambda_{i21}^{'^{*}}|\simeq 10^{-9}(m_{\tilde{l}}/100\, \rm GeV)^{2}$. 
In principle, we can avoid this bound considering {\it ad hoc} only
$\lambda'_{133},\lambda'_{233},\lambda'_{333}$
(only b-quark is involved).
This seems not justified by MSSM-like D-brane models: 
the three generations of $Q$ are attached to the same stacks,
as well as for the three of $L$, the three of $D^{c}$ and the three $U^{c}$.
However, coefficients coming from mixed disk correlators
are in general matrices with respect to flavors,
as mentioned in Section 2. 
So, possible hierarchies originated by mixed correlators
could also provide an intriguing motivation 
for direct channels at LHC 
avoiding "B,K-bounds"!
The most promising channel for LHC is associated to 
$\lambda'_{111}$. In our model a hierarchy 
of this with respect to $\lambda_{i12},...$ can be considered. 
A
resonant slepton productions 
in $pp\rightarrow eejj, ejj+\rm m.t.e$ \cite{Allanach1,Allanach2}
(m.t.e. is missing transverse energy)
can be envisaged, 
not necessary related to 
other bounds for other flavors. 
This channel can be also tested in $0\nu\beta\beta$-decays.
Finally, neutrino laboratories 
and astrophysics provide other interesting tests \cite{BDK}, compatible with
signals of opposite charged leptons $ee,\mu\mu,e\mu$ at LHC.

\subsection{R-violations with very light neutralini}
As remarked in \cite{SHIP1,SHIP2},
a light neutralino is 
excluded as a Dark Matter candidate
in R-parity preserving MSSM:
such a stable neutralino gives an excessive 
quantity of 
dark matter as a thermal relic. 
However, it is not ruled-out
in R-parity violating scenari.
In this case a range of masses
$0.7\, \rm eV<m_{\chi^{0}_{1}}<24\, \rm GeV$
can be considered: the excessive part of neutralini can decay to 
other particles through R-violating operators, with couplings  $O(10^{-6}\div 10^{-9})$.
On the other hand, non-thermal processes of dark matter 
production, such as Q-balls' decays in Affleck-Dine scenari \cite{AD1,AD2,AD3,Higaki:2014eda}, 
can strongly affect conventional calculations
on DM production (thermal production). 
In particular, Q-balls will not disrupted by R-violating operators 
for a sufficient time, if the involved RPV couplings are $O(10^{-6}\div 10^{-9})$.
This is  an interesting range for future researches in SHIP experiment \cite{SHIP1,SHIP2}.
For these motivations, a scenario in which 
$m_{\chi_{0}^{1}}<m_{B,D}$
is intriguing. 
In particular, a superpotential term as 
$\lambda'_{i21}L_{i}Q_{2}D^{c}_{1}$,
can lead to $D^{\pm} \rightarrow \chi_{0}^{1}+l^{\pm}_{i}$.
On the other hand, decays like $\chi_{0}^{1}\rightarrow \bar{K}_{S}^{0}\nu^{i},\bar{K}_{L}^{0}\nu^{i},K_{S}^{0}	\bar{\nu}^{i},K_{L}^{0}\bar{\nu}^{i}$ can be generated by the same operator. 
Analogous decays into charged Kaons can be considered, allowing $\lambda'_{i12}L_{i}Q_{1}D^{c}_{2}$.
These channels would be tested by SHiP experiment in next future \cite{SHIP1,SHIP2}.
In fact, as mentioned above, SHIP experiment could test new RPV couplings up to $10^{-9}$.
We would like to stress again that the introduction of one and only one operator,
among all possible R-violating ones, it is generically unnatural, 
while in our scenario is particularly simple to achieve as a dynamical R-parity breaking.
This theoretical argument enforces motivations in favor of these kinds of RPV researches. 

\subsection{EDMs}
Generically, a promising way
to detect indirect effects of R-parity violating 
trilinear terms is through Electric Dipole Moments (EDMs).
In fact, CP violating phases of $\mu',\lambda,\lambda',\lambda''$
have to contribute to EDMs of various baryons, nuclei, atoms and molecules. 
As extensively discussed in \cite{EDM}, 
Electric Dipole Moments (EDMs) of proton, deuteron, 
$He,Rn,Ra,Fr$, atoms, muons, and the R-correlation 
of neutron beta decay can constrain
RPV superpotential operators (\ref{RPV2})-(\ref{RPV4}).
In fact, the imaginary parts of $\lambda,\lambda'$ 
can  contribute with CP violating phases to 
EDMs, through Barr-Zee type two-loop contributions, 
and four-fermion interactions. 
Let us define, as done in \cite{EDM}, relevant combinations 
$$x_{1}=Im\left( \lambda_{311}\lambda_{322}^{*}\right)\,\,\,\,\,x_{2}=Im\left(\lambda_{211}\lambda_{233}^{*}\right)$$
$$x_{3}=Im\left(\lambda_{(i=2,3)11}\lambda_{(i=2,3)11}\right)\,\,\,\,x_{4}=Im\left( \lambda_{(i=2,3)11}\lambda_{(i=2,3)22}^{'*}\right)$$
$$x_{5}=Im \left( \lambda_{(i=2,3)11}\lambda_{(i=2,3)33}^{*'}\right)\,\,\,\,\,
x_{6}=Im\left( \lambda_{(i=1,2,3)22}\lambda_{(i=1,2,3)11}^{'*}\right)$$
$$x_{7}=Im \left( \lambda_{(i=1,2)33}\lambda_{(i=1,2)11}^{'*}\right)\,\,\,\,\,x_{8}=Im \left( \lambda_{(i=1,2,3)11}'\lambda_{(i=1,2,3)22}^{'*}\right)$$
$$x_{9}=Im\left(\lambda'_{(i=1,2,3)11}\lambda_{(i=1,2,3)33}^{'*} \right)\,\,\,\,
x_{10}=Im\left(\lambda'_{(i=1,2,3)22}\lambda_{(i=1,2,3)33}^{'*}\right)$$
These can be constrained by current available
EDM-data. 
For $TeV$-scale susy,
the upper bounds, obtained 
among all data, are  
$|x_{1}|<2\times 10^{-4}$
$|x_{2}|< 2\times 10^{-5}$, $|x_{3}|<2\times 10^{-8}$, 
$|x_{4}|<10^{-6}$, 
$|x_{5}|<7 \times 10^{-6}$,
$|x_{6}|<0.2$,
$|x_{7}|<2 \times 10^{-2}$,
$|x_{8}|< 7 \times 10^{-4}$, 
$|x_{9}|<3 \times 10^{-5}$, $|x_{10}|<2 \times 10^{-4}$.
In particular, $x_{1,2,3,4,5}$ are
constrained by ThO molecule, 
while $x_{6,7,8,9,10}$ by neutron dipole moment.

As regards R-correlation, this can constrain 
$Im\left(\lambda_{i11}\lambda_{i11}^{'*}\right)$ up to $10^{-10}$ for $Hg$ atom, if one assumes the dominance of only one 
"x". 

Let us comment these results in the light of D-brane models 
discussed above. Simpler cases are the ones 
in which only one $E2$-instanton generate only one bilinear or trilinear 
operator. In this cases, mixed combinations as $x_{4,5,6,7}$ are previewed 
to be zero. As a consequence, D-brane models for $\lambda$-couplings
seem to be disfavored with respect to $\lambda'$-models,
by EDMs data. 

\subsection{B-violating physics without proton decay}
Now, let us discuss the class of models 
with only $\lambda''$-terms. 
In this case, proton is not destabilized: lepton number is conserved,
avoiding $p\rightarrow  l^{+}\pi^{0},K^{+}\bar{\nu},...$. 
However, the strongest limits are placed by 
dinucleon decays $NN\rightarrow KK$,
$n-\bar{n}$ transitions 
and $n\rightarrow \varXi$. 
In particular, assuming $M_{susy}=1\, \rm  TeV$,
(or more precisely squarks masses around $1\,\rm TeV$), 
we can get approximately:  $|\lambda''_{112}|\sim |\lambda''_{11k}| \sim 10^{-6}$. Geometrically, 
this corresponds to an $E2$-instanton with large 3-cycles.  
 On the other hand, as discussed in the previous section, mixed disk amplitudes
are not necessary "democratic" with generations:
matrices with flavor indices emerge, and they can create
hierarchies between $\lambda''_{ijk}\sim C^{(2)}_{i}C^{(2)}_{j}C'^{(2)}_{k}$.
For each coefficient of matrices $C^{(2)}$ and $C'^{(2)}$, 
hierarchies of $10\div 10^{3}$ could be considered.
As a consequence, researches of direct signatures at LHC 
as B-violating decays can be interesting. 
In particular, processes like $\tilde{t}\rightarrow \bar{d}_{j}\bar{d}_{k}$
can be searched and well constrained, especially 
under the hypothesis 
of Long-Lived superparticles. 
For example, as shown in \cite{Liu}, 
for $\lambda''_{312}\sim 10^{-8}$
and $c\tau \sim 10^{-1}\div 1$, 
limits on $m_{\tilde{t}}$ arrives to $\simeq 900\, \rm GeV$.
Similar limits are obtained for $\lambda''_{333}$. 
Another possible decay channel in the case of a gluino LSP could be 
$\tilde{g}\rightarrow \tilde{q}q\rightarrow jjj$
where $U^{c}D^{c}D^{c}$ operator
split one squark into two quarks. 
In this case, regions of the parameters are
also more constrained than $\tilde{t}\rightarrow \bar{d}_{i}\bar{d}_{j}$
\cite{Liu}.
Alternatively, Higgsino three-body decays 
$\tilde{H} \rightarrow jjj$ can be also considered. 
In this case, limits are milder than the gluino-case \cite{Liu}.

 \subsection{Cosmological bounds on three linear superpotential terms.}
In this section, we would like to briefly remind
cosmological bounds on 
(\ref{RPV2})-(\ref{RPV3})-(\ref{RPV4}), from Baryogenesis and Leptogenesis. 
We also would like to mention possible ways-out.
This can be important for direct researches at LHC. 

Suppose to generate a Baryon or Lepton asymmetry 
in the primordial Universe before the electroweak phase transition
$E\gtrsim 100\, \rm GeV$. Under this quite generic hypothesis, 
we can put strong bounds on R-parity violating operators. 
In fact, they not conserve $B-L$. 
As a consequence, R-parity violating processes
wash-out $B-L$ component, while 
sphalerons wash out the $B+L$ one;
{\it i.e} any initial matter-antimatter asymmetry 
will be washed out! 
This leads to the upper bounds \cite{Rcosmology1,Rcosmology2,Davidson:1997mc}
\be{ub1}
\lambda_{ijk},\lambda'_{ijk},\lambda''_{ijk}<5\times 10^{-7}\sqrt{\frac{M_{SUSY}}{1\, \rm TeV}}
\ee
Clearly, these bounds have a possible way-out 
relaxing the initial assumption:
one can assume a Post-Sphaleron mechanism 
for baryogenesis.
However, here, 
we would like to suggest another 
possible idea as a sting-inspired way-out, alternative to Post-Sphalerons scenari:
it is possible that $\lambda'$, 
has grown during the cosmological time
as a "dynamical degree of freedom",
from a small value $\lambda'<10^{-7}$
up to a higher value reached in the present epoch $\lambda'>>10^{-7}$.  
In fact, in string-theories, 
all coupling constants depend on moduli, stabilized 
by non-perturbative effects like fluxes and instantons. 
But it is possible that $\lambda'$ can be stabilized 
not as a constant value but as a
 "solitonic solution" $\lambda'(t)$ with respect 
to the cosmological time $t$. 
The solitonic solution can connect two asymptotic branches 
$\lambda'(t<t_{early})<10^{-7}$ and $\lambda'(t>>t_{early})>>10^{-7}$.
This hypothesis can be constrained by BBN bounds, 
strongly depending on the particular working hypothesis
for the  
R-violating MSSM space of the parameters:
sparticles decays could ruin the right ratio 
of nuclei. 
Conservatively, one can assume $\lambda'(t<t_{BBN})\simeq \lambda'(t<t_{early})<10^{-7}$, 
in order to avoid any possible insidious
constraints from BBN. However this issue deserves deeper investigations 
beyond the purposes of this paper.

\section{Phenomenology for $M_{SUSY}>>1\, \rm TeV$}

In our class of effective models, we cannot predict  
the susy breaking scale.
It is undoubtable that MSSM is not in a good status after 
the first run of LHC. In TeV-scale susy, this favors R-parity violating MSSM with 
respect to R-preserving MSSM (more parameters).
However, 
the next run of LHC will definitely test   
both MSSM scenarios.
On the other hand, we cannot ignore the possibility that susy could 
be not linked to the hierarchy problem of the Higgs mass!
In fact supersymmetry could be important for other fundamental 
issues such as neutrino masses, baryon and lepton violations
and consistency of string theory
\footnote{Let us mention that in contest of non-local quantum field theories, 
supersymmetry seems an important element in order 
to cancel an infinite number of acausal divergences 
coming from F-terms \cite{Addazi:2015dxa}.  
In order to realize such a mechanism,  
susy can be broken at $\Lambda_{NL}$ (effective Non-locality scale),
supposed to be the Planck scale. 
On the other,
divergences of D-terms
remain uncured: susy is not a complete solution of the problem. 
In \cite{Addazi:2015ppa}, we also would like to mention that, recently, we have shown how the formation of a classical configuration in 
ultra-high energy scatterings could unitarize and causalize a non-local QFT. }
\footnote{In this case, an alternative dark matter candidate to neutralino 
 could be provided from a parallel intersecting D-branes' world. 
If the vev scale of this world is different form the vev scale of our ordinary one, 
a non-collisional dark halo, composed of dark atoms, can be obtained. 
A discussion of theoretical aspects 
and direct detection implications 
can be found in \cite{Addazi:2015cua}.}.
We would like to note that for $M_{susy}>>1\, \rm TeV$, R-parity violating MSSM remains still alive 
in phenomenology. For instance, assuming $\lambda\sim \lambda''\sim 1$
immediately we can put limits on $M_{susy}$ around the Planck scale, 
from proton decay limits.
In D-brane models, in which one $E2$-instanton generate 
only one R-parity violating bilinear or trilinear superpotential, 
a $M_{susy}>>1\, \rm TeV$ scenario can remain interesting. 
In these cases, proton decays will be avoided
if $\lambda, \lambda',\lambda''$ are not contemporary  
generated by the D-brane models, {\it i.e}
the correspondent three $E2$-instantons 
are not contemporary introduced.
In construction with one and only one 
among the possible bilinear and trilinear superpotentials,
limits on sparticles masses are much 
smaller than $M_{Pl}\simeq 10^{19}\, \rm GeV$.
As seen above, a situation in which $\lambda,\lambda',\lambda'' \sim 1$
is geometrically understood as $E2$-instantons wrapping 
3-cycles with small radii, on the $CY_{3}$. 
For example, let us consider 
the case shown in Fig.1-(b), corresponding to
a $\lambda''$-model. Supposing for example all $\lambda''\sim 1$, 
we can put an indirect bound on susy breaking scale 
from dinucleon decays, neutron-antineutron transitions 
and $n\rightarrow \varXi$, approximately corresponding to 
$M_{susy}\simeq 10^{2}\div 10^{3}\, \rm TeV$ 
(supposing squarks masses approximately equal to 
the susy breaking scale). 
The next generation of experiments
in neutron-antineutron oscillations 
promise to test the $10^{3}\, \rm TeV$
scale \cite{NNbar}. 
EDMs
 are other possible indirect test for this scenario.
 For example, the neutron electric dipole moment
 would be a good way to test PeV Scale Physics, in next future
 \cite{Aboubrahim:2015nza}.
 
\section{Higher order superpotential terms and further implications in $n-\bar{n}$ oscillations }
In previous sections, we have discussed 
R-parity violating bilinear and trilinear superpotentials. 
However, Exotic instantons can generate 
higher order superpotential terms without generating
at all bilinear and trilinear superpotential terms!
Examples are the ones considered 
in Fig.1-(a)-(b). The first one, as suggested 
in \cite{Ibanez2}, can directly generate a Weinberg operator
for neutrini masses $\mathcal{W}_{4}=H_{u}LH_{u}L/\Lambda_{4}$.
In fact, in the case of fig.1-(a), we can consider the same mixed disk amplitudes, 
but with an $E2$-instanton intersecting two times 
relevant D-brane stacks. 
Here, we suggest Fig.1-(b) can generate
$\mathcal{W}_{5}=(U^{c}D^{c}D^{c})^{2}/\Lambda_{5}^{3}$,
and consequently 
a Majorana mass for neutrons.
However, there is an important difference with respect to
previous cases: $\Lambda_{4,5}\sim e^{+S_{E2^{(IV)},E2^{(V)}}}M_{S}\geq M_{S}$. As a consequence, superpotentials like $\mathcal{W}_{4,5}$
are too much suppressed for phenomenology if, as usual, 
$M_{S}$ is only slightly smaller than the Planck scale. However, they can become 
interesting in low string scale scenari \cite{ADD1,AADD,ADD2,RS1,RS2}
\footnote{See \cite{Antonidias1,Antonidias2,Dimopoulos} for recent papers about the case of $1\div 10\, \rm TeV$ quantum gravity.}. For example we can envisage 
a $\Lambda_{5}\simeq M_{S}\simeq 10^{3}\, \rm TeV$ (small 3-cycles of $E2^{(V)}$-instanton). Also in this case, exotic instantons generate
a neutron-antineutron transition testable in the next future \cite{NNbar,Phillips:2014fgb}. 
As an alternative, one can generate analogous superpotentials
like $\mathcal{W}_{6}=QQD^{c}QQD^{c}/\Lambda_{6}^{3}$,
also leading to a neutron-antineutron transition.
In particular, from $\mathcal{W}_{5,6}$ we obtain 
the relevant New Physics (NP) scale
$\mathcal{M}_{5,6}^{5}=\Lambda_{5,6}^{3}m_{\tilde{g}}^{2}$,
where $m_{\tilde{g}}$ is the gaugino mass (gluino, zino or photino)
for operators 
$\mathcal{O}_{n\bar{n}}=(u_{R}d_{R}d_{R})^{2}/\mathcal{M}_{5}^{5}$ 
and $\mathcal{O}_{n\bar{n}}=(q_{L}q_{L}d_{R})^{2}/\mathcal{M}_{6}^{5}$.
As a consequence, neutron-antineutron bounds can be satisfied,
for example, for $m_{\tilde{g}}\simeq M_{S}\simeq 1000\, \rm TeV$, 
for $e^{+S_{E2}}\sim 1$ (small 3-cycles of $E2$-instanton on $CY_{3}$).
Alternatively, compatible with TeV-scale susy, 
$m_{\tilde{g}}\simeq 1\, \rm TeV$ and $\Lambda_{5,6}\simeq 10^{5}\, \rm TeV$ can also satisfy neutron-antineutron bounds
\footnote{ Neutron-Antineutron oscillation could be also a probe for CPT symmetry \cite{Okun} and new fifth force interactions \cite{AddaziIFAE}.}. 

\section{Conclusions and remarks}
In this paper, we have shown how 
R-parity can be dynamically broken by Exotic Instantons 
in a simple, calculable and controllable way, 
in a class of D-brane models. We have discussed 
explicit examples of intersecting D-branes,
generating a RV bilinear or trilinear terms in the superpotential. 
We have stressed how one $E2$-instanton generates 
one and only one bilinear or trilinear term, 
without generating the other ones.
In this sense, a dynamical breaking of R-parity is
radically different with respect to an explicit one.
In fact, in explicitly R-violating models,
in principle one has to consider all possible 
R-violating operators in the superpotential,
{\it i.e.} a strong fine-tuning is necessary 
in order to avoid proton decay.
This unnatural situation is avoided in an elegant way 
in D-brane models. 
Another important feature of Exotic instantons 
is that they are not necessary "democratic"
with flavors, depending on the particular 
topology of the mixed disk amplitude. 
This can enforce reasons for direct tests at LHC,
 avoiding a lot of indirect bounds 
from meson physics, FCNCs and so on.
In addition, we have also commented the 
possible generation of higher dimensional 
operators in the superpotential, 
dynamically breaking R-parity, 
without generating bilinear or trilinear ones.
In several different scenarios, 
we have discussed phenomenological implications
in neutrino physics, neutron physics, EDMs, Dark Matter and LHC.
We conclude that string theory provides
powerful tools for phenomenology
of Baryon and Lepton number violations:
 exotic instantons could be key elements for the 
understanding of many aspects of fundamental physics. 

\vspace{1cm} 

{\large \bf Acknowledgments} 
\vspace{3mm}

I would like to thank Massimo Bianchi for illuminating discussions
and Ryan Griffiths for useful remarks. 
I also would like to thanks organizers of Eurostring 2015, in Cambridge (UK) for the hospitality, during the preparation of this paper. 
My work was supported in part by the MIUR research
grant "Theoretical Astroparticle Physics" PRIN 2012CPPYP7
and SdC Progetto Speciale Multiasse La Societ\'a  della Conoscenza in
Abruzzo, PO FSE Abruzzo 2007 - 2013. 



\begin{thebibliography}{99}

\bibitem{R1}
L. J. Hall and M. Suzuki, 
Nucl.Phys., vol. B231, p. 419, 1984.

\bibitem{R2}
G. G. Ross and J. Valle, 
Phys.Lett., vol. B151, p. 375, 1985.

\bibitem{R3}
V. D. Barger, G. Giudice, and T. Han, 
 Phys.Rev., vol. D40, p. 2987, 1989.

\bibitem{R5}
H. K. Dreiner, 
Adv.Ser.Direct.High Energy Phys., vol. 21, pp. 565Ð583, 2010.

\bibitem{R6}
G. Bhattacharyya, 
ÒA Brief Review of R-Parity Violating Couplings,Ó In *Tegernsee 1997,Beyond the desert 1997* 194-201, 1997.


\bibitem{R7}
R. Barbier, C. Berat, M. Besancon, M. Chemtob, A. Deandrea, E. Dudas, P. Fayet and
S. Lavignac et al., Phys. Rept. 420, 1 (2005).

\bibitem{R8}
E. Nikolidakis and C. Smith, 
Phys.Rev., vol. D77, p. 015021, 2008.

\bibitem{R9}
C. Csaki, Y. Grossman, and B. Heidenreich, 
Phys.Rev., vol. D85, p. 095009, 2012.

\bibitem{R10}
G. Krnjaic and D. Stolarski, 
JHEP, vol. 1304, p. 064, 2013.

\bibitem{R11}
R. Franceschini and R. Mohapatra, 
JHEP, vol. 1304, p. 098, 2013.

\bibitem{R12}
C. Csaki and B. Heidenreich,
Phys.Rev.,vol. D88, p. 055023, 2013.

\bibitem{R13}
G. Krnjaic and Y. Tsai, 
JHEP, vol. 1403, p. 104, 2014.

\bibitem{R14}
C. Brust, A. Katz, and R. Sundrum, 
JHEP, vol. 1208, p. 059, 2012.

\bibitem{R15}
P. W. Graham, D. E. Kaplan, S. Rajendran, and P. Saraswat, 
JHEP, vol. 1207, p. 149, 2012.

\bibitem{R16}
P. Fileviez Perez and S. Spinner, 
JHEP, vol. 1204, p. 118, 2012.

\bibitem{R17}
Z. Han, A. Katz, M. Son, and B. Tweedie, 
Phys.Rev., vol. D87, no. 7, p. 075003, 2013.

\bibitem{R18}
J. Berger, C. Csaki, Y. Grossman, and B. Heidenreich, 
Eur.Phys.J., vol. C73, no. 4, p. 2408, 2013.

\bibitem{R19}
R. Franceschini and R. Torre, 
Eur.Phys.J., vol. C73, p. 2422, 2013.

\bibitem{R20}
J. T. Ruderman, T. R. Slatyer, and N. Weiner, 
JHEP,vol. 1309, p. 094, 2013.


\bibitem{Csaki1}
C. Csaki, E. Kuflik, and T. Volansky, 
 Phys.Rev.Lett.,
vol. 112, p. 131801, 2014.


\bibitem{Csaki2}  
  C.~Csaki, E.~Kuflik, O.~Slone and T.~Volansky,
  arXiv:1502.03096 [hep-ph].
  
\bibitem{Perez1}
  V.~Barger, P.~Fileviez Perez and S.~Spinner,
  Phys.\ Rev.\ Lett.\  {\bf 102} (2009) 181802
  [arXiv:0812.3661 [hep-ph]].
  
\bibitem{Perez2}
  P.~Fileviez Perez and S.~Spinner,
  Phys.\ Lett.\ B {\bf 673} (2009) 251
  [arXiv:0811.3424 [hep-ph]].
  
\bibitem{Perez3}
  P.~Fileviez Perez,
  Int.\ J.\ Mod.\ Phys.\ A {\bf 28} (2013) 1330024
  [arXiv:1305.6935 [hep-ph]].
  
\bibitem{Ibanez1}
  L.~E.~Ibanez and A.~M.~Uranga,
  JHEP {\bf 0703} (2007) 052
  [hep-th/0609213].
  
\bibitem{Ibanez2}
  L.~E.~Ibanez, A.~N.~Schellekens and A.~M.~Uranga,
  JHEP {\bf 0706} (2007) 011
  [arXiv:0704.1079 [hep-th]].
  
  \bibitem{Ibanez3}
  L. E. Ibanez and A. M. Uranga, 
  ÒString theory and particle physics: An introduction
to string phenomenology,Ó Cambridge, UK: Univ. Pr. (2012) 673 p.
  
\bibitem{Blu1}
  R.~Blumenhagen, M.~Cvetic, D.~Lust, R.~Richter and T.~Weigand,
  Phys.\ Rev.\ Lett.\  {\bf 100} (2008) 061602
  [arXiv:0707.1871 [hep-th]].
  
\bibitem{Blu2}
  R.~Blumenhagen, M.~Cvetic and T.~Weigand,
  Nucl.\ Phys.\ B {\bf 771} (2007) 113
  [hep-th/0609191].
  
\bibitem{Florea}
B. Florea, S. Kachru, J. McGreevy, N. Saulina, 
JHEP 0705 (2007) 024. [hep-th/0610003].

\bibitem{Cvetic1}
  M.~Cvetic, J.~Halverson, P.~Langacker and R.~Richter,
  JHEP {\bf 1010} (2010) 094
  [arXiv:1001.3148 [hep-th]].
  
 \bibitem{Cvetic2} 
 M. Cvetic, T. Weigand, 
 Phys. Rev. Lett. 100 (2008) 251601. [arXiv:0711.0209 [hep-th]].
  
\bibitem{Cvetic3}
M. Cvetic, R. Richter, T.Weigand, Ò
Phys. Rev. D76 (2007) 086002. [hep-th/0703028].

\bibitem{Addazi1}
  A.~Addazi and M.~Bianchi,
  JHEP {\bf 1412} (2014) 089
  [arXiv:1407.2897 [hep-ph]].

\bibitem{Addazi2}
   A.~Addazi,
  JHEP {\bf 1504} (2015) 153
  [arXiv:1501.04660 [hep-ph]].
  
\bibitem{Addazi3}
 A.~Addazi and M.~Bianchi,
  JHEP {\bf 1507} (2015) 144
  [arXiv:1502.01531 [hep-ph]].
  
\bibitem{Addazi4}
  A.~Addazi and M.~Bianchi,
  JHEP {\bf 1506} (2015) 012
  [arXiv:1502.08041 [hep-ph]].
    
\bibitem{Addazi5}
  A.~Addazi,
  arXiv:1504.06799 [hep-ph].
  
\bibitem{Addazi7}
  A.~Addazi,
  arXiv:1505.02080 [hep-ph].
  
\bibitem{Addazi:2015goa}
  A.~Addazi,
  arXiv:1506.06351 [hep-ph].
 
  
\bibitem{Koba1}
  Y.~Hamada, T.~Kobayashi and S.~Uemura,
  JHEP {\bf 1405} (2014) 116
  [arXiv:1402.2052 [hep-th]].
  
    
\bibitem{Koba2}
  H.~Abe, T.~Kobayashi, Y.~Tatsuta and S.~Uemura,
  arXiv:1502.03582 [hep-ph].
  
    \bibitem{Parsa1}
   H.~Ghorbani, D.~Musso and A.~Lerda,
  JHEP {\bf 1103} (2011) 052
  [arXiv:1012.1122 [hep-th]].
  
  \bibitem{Parsa2}
  H.~Ghorbani and D.~Musso,
  JHEP {\bf 1112} (2011) 070
  [arXiv:1111.0842 [hep-th]].
  
  \bibitem{Parsa3}
   H.~Ghorbani,
  JHEP {\bf 1312} (2013) 041
  [arXiv:1306.1487 [hep-th]].
  
\bibitem{Addazi:2014mga}
  A.~Addazi and S.~Capozziello,
 International Journal of Theoretical Physics, 1-12.   arXiv:1407.4840 [gr-qc].
    
    
\bibitem{DMSSM1}  
  L. E. Ibanez, F. Marchesano and R. Rabadan, 
  JHEP 0111, 002 (2001), hep-th/0105155.

\bibitem{DMSSM2}  
  R. Blumenhagen, B. Kors, D. Lust and T. Ott, 
  Nucl. Phys. B {\bf 616}, 3 (2001), hep-th/0107138.
  
  \bibitem{DMSSM3}
  M. Cvetic, G. Shiu and A. M. Uranga, 
  Phys. Rev. Lett. {\bf 87}, 201801 (2001), hep-th/0107143. 
Nucl. Phys. B {\bf 615}, 3 (2001), hep-th/0107166.
  
  \bibitem{DMSSM4}
  D. Bailin, G. V. Kraniotis and A. Love,
Phys. Lett. B {\bf 530}, 202 (2002), hep-th/0108131. 
hep-th/0210219.
  
  \bibitem{DMSSM5}
  D. Cremades, L. E. Ibanez and F. Marchesano, 
  JHEP 0207, 009 (2002), hep-th/0201205.
  
  \bibitem{DMSSM6}
  D. Cremades, L. E. Ibanez and F. Marchesano, 
  JHEP 0207, 022 (2002), hep-th/0203160.
  
\bibitem{DMSSM8}
  L.~E.~Ibanez,
  hep-ph/0109082.
  
\bibitem{DMSSM9}
  R.~Blumenhagen, V.~Braun, B.~Kors and D.~Lust,
  hep-th/0210083.
  
 \bibitem{DMSSM10} 
D. Cremades, L. E. Ibanez and F. Marchesano,
 hep-ph/0212048.
 
\bibitem{DMSSM11}
  D.~Lust,
  Class.\ Quant.\ Grav.\  {\bf 21} (2004) S1399
  [hep-th/0401156].
 
 \bibitem{DMSSM12}
G. Aldazabal, L. E. Ibanez, F. Quevedo, and A. M. Uranga, 
JHEP 08 (2000) 002, arXiv:hep-th/0005067.

\bibitem{DMSSM13}
D. Berenstein, V. Jejjala, and R. G. Leigh, 
Phys.Rev.Lett. 88 (2002) 071602, arXiv:hep-ph/0105042 [hep-ph].

\bibitem{DMSSM14}
G. Aldazabal, L. E. Ibanez, and F. Quevedo, 
JHEP 0002 (2000) 015, arXiv:hep-ph/0001083 [hep-ph].

\bibitem{DMSSM15}
  J. Cascales, M. Garcia del Moral, F. Quevedo, and A. Uranga, 
  JHEP 0402 (2004) 031, arXiv:hep-th/0312051 [hep-th].
  
    \bibitem{Jim1}
  M.~Cvetic, J.~Halverson and R.~Richter,
  JHEP {\bf 1007} (2010) 005
  [arXiv:0909.4292 [hep-th]].
  
  \bibitem{Jim2}
  M.~Cvetic, J.~Halverson and P.~Langacker,
  JHEP {\bf 1111} (2011) 058
  [arXiv:1108.5187 [hep-ph]].

  \bibitem{Jim3}
  J.~Halverson,
  Phys.\ Rev.\ Lett.\  {\bf 111} (2013) 26,  261601
  [arXiv:1310.1091 [hep-th]].
  
  \bibitem{Jim4}
   M.~Cvetic, J.~Halverson and R.~Richter,
  arXiv:0910.2239 [hep-th].
  
\bibitem{GUT1}
  B.~Bajc and L.~Di Luzio,
  arXiv:1502.07968 [hep-ph].
  
\bibitem{Valle1}
  S.~M.~Boucenna, S.~Morisi and J.~W.~F.~Valle,
  Phys.\ Rev.\ D {\bf 90} (2014) 1,  013005
  [arXiv:1405.2332 [hep-ph]].
  
\bibitem{Valle2}
  S.~M.~Boucenna, J.~W.~F.~Valle and A.~Vicente,
  arXiv:1502.07546 [hep-ph].
  
\bibitem{Valle3}
  S.~M.~Boucenna, R.~M.~Fonseca, F.~Gonzalez-Canales and J.~W.~F.~Valle,
  Phys.\ Rev.\ D {\bf 91} (2015) 3,  031702
  [arXiv:1411.0566 [hep-ph]].

  
\bibitem{GCS1}
  P.~Anastasopoulos, M.~Bianchi, E.~Dudas and E.~Kiritsis,
  JHEP {\bf 0611} (2006) 057
  [hep-th/0605225].
  
\bibitem{GCS2}
J. De Rydt, J. Rosseel, T.T. Schmidt, A. Van Proeyen and M. Zagermann, 
Class. Quant. Grav. {\bf 24} (2007) 5201 [arXiv:0705.4216] [INSPIRE].

\bibitem{GCS3} 
D. Feldman, Z. Liu and P. Nath, 
Phys. Rev. D {\bf 75} (2007) 115001 [hep-ph/0702123] [INSPIRE].

\bibitem{GCS4}
 D. Feldman, Z. Liu and P. Nath, 
 AIP Conf. Proc. {\bf 939} (2007) 50 [arXiv:0705.2924] [INSPIRE].

\bibitem{GCS5}
B. K$\ddot{o}$rs and P. Nath, 
Phys. Lett. {\bf B 586}
(2004) 366 [hep-ph/0402047] [INSPIRE].

\bibitem{GCS6}
 B. K$\ddot{o}$rs and P. Nath, 
 JHEP {\bf 12} (2004) 005 [hep-ph/0406167] [INSPIRE].
 
\bibitem{GCS7}
  B. K$\ddot{o}$rs and P. Nath, 
hep-ph/0411406 [INSPIRE].

 \bibitem{GCS8}
 B. K$\ddot{o}$rs and P. Nath, 
 JHEP {\bf 07} (2005) 069 [hep-ph/0503208] [INSPIRE].
  
    
  \bibitem{GCS9}
  P. Anastasopoulos, F. Fucito, A. Lionetto, G. Pradisi, A. Racioppi and Y.S. Stanev,
  Phys. Rev. D {\bf 78} (2008) 085014 [arXiv:0804.1156] [INSPIRE].
  
  \bibitem{GCS10}
  C. Corian\'o, N. Irges and E. Kiritsis, 
  Nucl. Phys. B {\bf 746} (2006) 77 [hep-ph/0510332] [INSPIRE].
  
 \bibitem{GCS11}
 M. Bianchi and E. Kiritsis, Nucl. Phys. B {\bf 782} (2007) 26 [hep-th/0702015]. 
  
  \bibitem{Sagnotti1}
A. Sagnotti, 
Phys. Lett. {\bf B294} (1992) 196 [hep-th/9210127].

\bibitem{Sagnotti2}
C. Angelantonj and A. Sagnotti, 
Phys. Rept. 1 [(Erratum-ibid.) 339] arXiv:hep-th/0204089.

\bibitem{Sagnotti3}
G.~Pradisi and A.~Sagnotti,
Phys.\ Lett.\ B {\bf 216} (1989) 59.


\bibitem{Sagnotti6}
D. Fioravanti, G. Pradisi and A. Sagnotti, 
Phys. Lett. B{\bf 321} (1994) 349 [arXiv:hep-th/9311183].

\bibitem{Sagnotti7}
 A. Sagnotti, 
 in *Palaiseau 1995, Susy 95* 473-484 [hep-th/9509080].

\bibitem{Sagnotti8}
A. Sagnotti, 
Nucl. Phys. Proc. Suppl. {\bf 56B} (1997) 332 [arXiv:hep-th/9702093].

  
  
  
   \bibitem{Bianchi:1990yu}
  M.~Bianchi and A.~Sagnotti,
  Phys.\ Lett.\ B {\bf 247} (1990) 517.

\bibitem{Bianchi:1990tb}
  M.~Bianchi and A.~Sagnotti,
  Nucl.\ Phys.\ B {\bf 361} (1991) 519.

\bibitem{Bianchi:1991eu}
  M.~Bianchi, G.~Pradisi and A.~Sagnotti,
  Nucl.\ Phys.\ B {\bf 376} (1992) 365.

\bibitem{sessantatre}
A. M. Uranga, Nucl. Phys. B {\bf 598}, 225 (2001) [hep-th/0011048].

\bibitem{sessantaquattro}
G. Aldazabal, S. Franco, L. E. Ibanez, R. Rabadan and A. M. Uranga, J. Math. Phys. {\bf 42},
3103 (2001) [hep-th/0011073].

\bibitem{MBJFM}
  M.~Bianchi and J.~F.~Morales,
  JHEP {\bf 0003} (2000) 030
  [hep-th/0002149].

\bibitem{Angelantonj:1996uy}
  C.~Angelantonj, M.~Bianchi, G.~Pradisi, A.~Sagnotti and Y.~.S.~Stanev,
  Phys.\ Lett.\ B {\bf 385} (1996) 96
  [hep-th/9606169].

\bibitem{Angelantonj:1996mw}
  C.~Angelantonj, M.~Bianchi, G.~Pradisi, A.~Sagnotti and Y.~S.~Stanev,
  Phys.\ Lett.\ B {\bf 387} (1996) 743
  [hep-th/9607229].
  
  
\bibitem{17}
M.Hirsch, A.M. Diaz, W.Porod, J.C. Romao and J.W.F Valle, Phys. Rev. D {\bf 62} 113008 [Erratum-ibid.
2002 D {\bf 65} 119901] (Preprint arXiv:hep-ph/0004115).

\bibitem{18}
M.A. Diaz, M.Hirsch, W.Porod, J.C. Romao and J.W.F. Valle Phys. Rev. D {\bf 68} 013009 [Erratum-ibid.
2005 D {\bf 71} 059904] (Preprint arXiv:hep-ph/0302021). 
  
\bibitem{20}
 F. de Campos, O.J.P. Eboli, M.B. Magro, W. Porod, D.Restrepo, S.P.Das, M.Hirsch, J.W.F.Valle,
Phys. Rev. D {\bf 86} 075001 (Preprint arXiv:1206.3605 [hep-ph]).

\bibitem{30a}
 F. De Campos, M.A. Diaz, O.J.P. Eboli, M.B. Magro and P.G.Mercadante Nucl. Phys. B {\bf 623} 47 (Preprint
hep-ph/0110049);
  
\bibitem{30b}
F. De Campos, M.A. Diaz, O.J.P. Eboli, R.A. Lineros, M.B. Magro and P.G. Mercadante Phys. Rev. D
{\bf 71} 055008 (Preprint hep-ph/0409043)

\bibitem{30c}
F. De Campos, M.A. Diaz, O.J.P. Eboli, M.B. Magro, W.Porod and S.Skadhauge, Phys. Rev. D {\bf 77} 115025 (Preprint arXiv:0803.4405 [hep-ph]).
  
 \bibitem{M1}
V.A. Mitsou, arXiv:1502.07997v1 [hep-ph] 27 Feb 2015. 
  
\bibitem{Cheriguene:2014bxa}
  A.~Cheriguene, S.~Liebler and W.~Porod,
  Phys.\ Rev.\ D {\bf 90} (2014) 5,  055012
  [arXiv:1406.7837 [hep-ph]].
  
\bibitem{Liu:2015bma}
  Z.~Liu and B.~Tweedie,
  arXiv:1503.05923 [hep-ph].
\bibitem{Gravitino1}
S. T. Butler and C. A. Pearson, 
Phys. Rev. {\bf 129} (1963) 836.

\bibitem{Gravitino2}
A. Schwarzschild and C.Zupancic
Phys. Rev. {\bf 129} (1963) 854.

\bibitem{Gravitino3}
P. Chardonnet, J. Orlo and P. Salati, 
Phys. Lett. B {\bf 409} (1997) 313 [arXiv:astro-ph/9705110].

\bibitem{Gravitino4}
F. Takayama and M. Yamaguchi, 
Phys.Lett. B {\bf 485} (2000) 388 [arXiv:hep-ph/0005214].

\bibitem{Gravitino5}
G. Moreau and M. Chemtob, 
Phys. Rev. D {\bf 65} (2002) 024033 [arXiv:hep-ph/0107286].

\bibitem{Gravitino6}
W.Buchmuller, L.Covi, K.Hamaguchi, A.Ibarra and T.Yanagida, 
JHEP 0703 (2007) 037 [arXiv:hepph/0702184].

\bibitem{Gravitino7}
 N.-E. Bomark, S. Lola, P. Osland and A. R. Raklev, 
 Phys. Lett. B {\bf 677} (2009) 62 [arXiv:0811.2969 [hep-ph]].

\bibitem{Gravitino8}
L. Covi, M. Grefe, A. Ibarra and D. Tran, 
JCAP 0901 (2009) 029 [arXiv:0809.5030 [hep-ph]].

\bibitem{Gravitino9}
L. A. Dal and A. R. Raklev, 
Phys. Rev. D {\bf 89} (2014) 103504 [arXiv:1402.6259 [hepph]].


\bibitem{Gravitino10}
  M.~Grefe and T.~Delahaye,
  arXiv:1503.01101 [hep-ph].
  
  \bibitem{A1}
  M. Hirsch, H. V. Klapdor-Kleingrothaus and S. G. Kovalenko, Phys. Rev. D {\bf 53}, 1329 (1996)
[arXiv:hep-ph/9502385].
  
  \bibitem{A2}
  A. Faessler, S. Kovalenko and F. Simkovic, Phys. Rev. D {\bf 58} (1998) 115004 [arXiv:hep-ph/9803253].
  
  \bibitem{A3}
  M. Hirsch, H. V. Klapdor-Kleingrothaus and S. G. Kovalenko, Phys. Lett. B {\bf 372}, 181 (1996)
[Erratum-ibid. B {\bf 381}, 488 (1996)] [arXiv:hep-ph/9512237].
  
  \bibitem{A4}
  H. P$\ddot{a}$s, M. Hirsch and H. V. Klapdor-Kleingrothaus, Phys. Lett. B {\bf 459}, 450 (1999)
[arXiv:hep-ph/9810382]. 
  
\bibitem{A5}
  A.~Faessler, T.~Gutsche, S.~Kovalenko and F.~Simkovic,
  Phys.\ Rev.\ D {\bf 77} (2008) 113012
  [arXiv:0710.3199 [hep-ph]].
  
\bibitem{Al1}
  B.~C.~Allanach, C.~H.~Kom and H.~P$\ddot{a}$s,
  Phys.\ Rev.\ Lett.\  {\bf 103} (2009) 091801
  [arXiv:0902.4697 [hep-ph]].
  
\bibitem{Al2}
  B.~C.~Allanach, C.~H.~Kom and H.~P$\ddot{a}$s,
  JHEP {\bf 0910} (2009) 026
  [arXiv:0903.0347 [hep-ph]].
  
  \bibitem{KS}
A. Kundu and J. P. Saha, Phys. Rev. D{\bf 70}, 096002 (2004) [hep-ph/0403154].
  
    \bibitem{Allanach1}
  B.~Allanach, A.~R.~Raklev and A.~Kvellestad,
  arXiv:1409.3532 [hep-ph].
  
  \bibitem{Allanach2}
   B.~C.~Allanach, S.~Biswas, S.~Mondal and M.~Mitra,
  Phys.\ Rev.\ D {\bf 91} (2015) 1,  015011
  [arXiv:1410.5947 [hep-ph]].
  
  \bibitem{BDK}
  R.~Bose, A.~Datta, A.~Kundu and S.~Poddar,
  Phys.\ Rev.\ D {\bf 90} (2014) 3,  035007
  [arXiv:1405.1282 [hep-ph]].
  
  \bibitem{SHIP1}
  S.~Alekhin, W.~Altmannshofer, T.~Asaka, B.~Batell, F.~Bezrukov, K.~Bondarenko, A.~Boyarsky and N.~Craig {\it et al.},
  arXiv:1504.04855 [hep-ph].
  
  \bibitem{SHIP2}
  S.~Alekhin, W.~Altmannshofer, T.~Asaka, B.~Batell, F.~Bezrukov, K.~Bondarenko, A.~Boyarsky and N.~Craig {\it et al.},
  arXiv:1504.04855 [hep-ph].
  
  \bibitem{AD1}
  I. Affleck and M. Dine, Nucl.Phys. {\bf B249}, 361 (1985).
  
  \bibitem{AD2}
  M. Dine, L. Randall, and S. D. Thomas, Nucl.Phys. {\bf B458}, 291 (1996), hep-ph/9507453.
  
  \bibitem{AD3}
  M. Dine and A. Kusenko, Rev.Mod.Phys. {\bf 76}, 1 (2003), hep-ph/0303065.
  
\bibitem{Higaki:2014eda}
  T.~Higaki, K.~Nakayama, K.~Saikawa, T.~Takahashi and M.~Yamaguchi,
  Phys.\ Rev.\ D {\bf 90} (2014) 4,  045001
  [arXiv:1404.5796 [hep-ph]].
  
  
\bibitem{EDM}
  N.~Yamanaka, T.~Sato and T.~Kubota,
  JHEP {\bf 1412} (2014) 110
  [arXiv:1406.3713 [hep-ph]].
  
    \bibitem{NNbar}
  D.~G.~Phillips, II, W.~M.~Snow, K.~Babu, S.~Banerjee, D.~V.~Baxter, Z.~Berezhiani, M.~Bergevin and S.~Bhattacharya {\it et al.},
  [arXiv:1410.1100 [hep-ex]].
  
    \bibitem{Aboubrahim:2015nza}
  A.~Aboubrahim, T.~Ibrahim and P.~Nath,
  arXiv:1503.06850 [hep-ph].
   
\bibitem{Liu}
    Z.~Liu and B.~Tweedie,
  arXiv:1503.05923 [hep-ph].
  
   \bibitem{Rcosmology1}
  B.A. Campbell, S.Davidson, J. Ellis and K.A. Olive, Phys. Lett. {\bf B 256} (1991) 457, Astroparticle Phys. {\bf 1} (1992) 77.
  
  \bibitem{Rcosmology2}
  H. Dreiner and G.G. Ross, Nucl. Phys. {\bf B410} (1993) 188.
  
\bibitem{Davidson:1997mc}
  S.~Davidson and J.~R.~Ellis,
  Phys.\ Rev.\ D {\bf 56} (1997) 4182
  [hep-ph/9702247].
  
\bibitem{Addazi:2015dxa}
  A.~Addazi and G.~Esposito,
  arXiv:1502.01471 [hep-th].
  
\bibitem{Addazi:2015ppa}
  A.~Addazi,
  arXiv:1505.07357 [hep-th].
  
\bibitem{Addazi:2015cua}
  A.~Addazi, Z.~Berezhiani, R.~Bernabei, P.~Belli, F.~Cappella, R.~Cerulli and A.~Incicchitti,
  Eur.\ Phys.\ J.\ C {\bf 75} (2015) 8,  400
  [arXiv:1507.04317 [hep-ex]].
  

\bibitem{ADD1}
N. Arkani-Hamed, S. Dimopoulos, G. Dvali (1998). 
Physics Letters {\bf B429} (3Ð4): 263Ð272. [arXiv:hep-ph/9803315]

\bibitem{AADD}
 I. Antoniadis, N. Arkani-Hamed, S. Dimopoulos, G. Dvali (1998).
 Physics Letters {\bf B436} (3Ð4): 257Ð263. [arXiv:hep-ph/9804398]. 
 
 \bibitem{ADD2}
N. Arkani-Hamed, S. Dimopoulos, G. Dvali (1999). 
Physical Review {\bf D59} (8): 086004. [arXiv:hep-ph/9807344]
  
\bibitem{RS1}
  L.~Randall and R.~Sundrum,
  Phys.\ Rev.\ Lett.\  {\bf 83} (1999) 3370
  [hep-ph/9905221].

\bibitem{RS2}
  L.~Randall and R.~Sundrum,
  Nucl.\ Phys.\ B {\bf 557} (1999) 79
  [hep-th/9810155].

\bibitem{Antonidias1}
  I.~Antoniadis,
  Int.\ J.\ Mod.\ Phys.\ A {\bf 29} (2014) 1444001.

\bibitem{Antonidias2}
  L.~A.~Anchordoqui, I.~Antoniadis, D.~C.~Dai, W.~Z.~Feng, H.~Goldberg, X.~Huang, D.~Lust and D.~Stojkovic {\it et al.},
  Phys.\ Rev.\ D {\bf 90} (2014) 6,  066013
  [arXiv:1407.8120 [hep-ph]].

\bibitem{Dimopoulos}
  S.~Dimopoulos, K.~Howe, J.~March-Russell and J.~Scoville,
  arXiv:1412.0805 [hep-ph].
  
 \bibitem{NNbar}
M. Baldo-Ceolin {\it et al.},
Z. Phys. C {\bf 63} (1994) 409.  

\bibitem{Phillips:2014fgb}
  D.~G.~Phillips, II, W.~M.~Snow, K.~Babu, S.~Banerjee, D.~V.~Baxter, Z.~Berezhiani, M.~Bergevin and S.~Bhattacharya {\it et al.},
  [arXiv:1410.1100 [hep-ex]].
  
\bibitem{Okun}  
Yu. Abov, F. Dzheparov, L.B. Okun, JETP Lett. {\bf 39}, 493 (1984). 
  
 \bibitem{AddaziIFAE}
 A.~Addazi,
  Nuovo Cim.\ C {\bf 038} (2015) 01,  21.


\end{thebibliography}
\end{document}